\documentclass[twocolumn,pre,floatfix,showpacs]{revtex4}

\usepackage{epsfig}
\usepackage{times}

\usepackage{ifthen}

\def\be{\begin{equation}}
\def\ee{\end{equation}}
\def\ep{\epsilon}
\def\te{\tilde{\epsilon}}

\begin{document}
\bibliographystyle{revtex}

\title{Electrokinetic behavior of two touching inhomogeneous biological
cells and colloidal particles: Effects of multipolar interactions}

\author{J. P. Huang}
\affiliation{Department of Physics, The Chinese University of Hong Kong,
 Shatin, NT, Hong Kong, and\\
Biophysics and Statistical Mechanics Group,
Laboratory of Computational Engineering, Helsinki University of Technology,
 P.\,O. Box 9203, FIN-02015 HUT, Finland}

\author{Mikko Karttunen}
\affiliation{Biophysics and Statistical Mechanics Group,
Laboratory of Computational Engineering, Helsinki University of Technology,
 P.\,O. Box 9203, FIN-02015 HUT, Finland}

\author{K. W. Yu and L. Dong}
\affiliation{
Department of Physics, The Chinese University of Hong Kong,
Shatin, NT, Hong Kong}

\author{G. Q. Gu}
\affiliation{Department of Physics, The Chinese University of Hong Kong,
 Shatin, NT, Hong Kong, and College of Information Science and Technology, East China Normal
University, Shanghai 200 062, China}

\begin{abstract}

We present a theory to investigate electro-kinetic behavior,
namely, electrorotation and dielectrophoresis under alternating
current (AC) applied fields for a pair of touching inhomogeneous
colloidal particles and biological cells. These inhomogeneous
particles are treated as graded ones with physically motivated
model dielectric and conductivity profiles. The mutual
polarization interaction between the particles yields a change in
their respective dipole moments, and hence in the AC
electrokinetic spectra. The multipolar interactions between
polarized particles are accurately captured by the multiple images
method. In the point-dipole limit, our theory reproduces the known
results. We find that the multipolar interactions as well as the
spatial fluctuations inside the particles can affect the AC
electrokinetic spectra significantly.
\end{abstract}

%%%%%%%%%%%%
% 82.70.-y Disperse systems; complex fluids
% 77.22.Gm Dielectric loss and relaxation
% 61.20.Qg Structure of associated liquids: electrolytes, molten salts, etc.
% 77.84.Nh Liquids, emulsions, and suspensions; liquid crystals 
% 77.22.-d Dielectric properties of solids and liquids
% 77.22.Ej Polarization and depolarization
% 77.84.Lf Composite materials
% 42.79.Ry Gradient-index (GRIN) devices 
%%%%%%%%%%%%%%%


\maketitle

\section{Introduction}

Identification and analysis of cell populations and (micro)biological
particles are essential in many practical applications ranging from
cancer research to chemical analysis of environmental pollutants.
During the past decade, alternating current (AC) electrokinetic phenomena,
and in particular electrorotation (ER) and dielectrophoresis (DEP),  have received
much attention in this respect, especially in  micromanipulation
and separation of submicron size
particles~\cite{Duan,Hug1,Mor1,Chou,ratanachoo02a,Zim1,Fuhr1,yang:1999a,YH2,MH1,JPCM1,Gascoyne:2002a}.
In addition to biological and environmental applications, AC electrokinetic
phenomena have been suggested as possible mechanisms for
nanomotors~\cite{hughes00a,Hughes:2002b}.

Both dielectrophoresis and electrorotation are based on
dielectric properties of  particles.
These properties depend heavily on the nature of the surface,
e.g., size, shape, and charge density. For example, since the composition
and shape of cancer cells differ from those of healthy cells,
these difference are reflected in their characteristic dielectric
properties which can be exploited in identifying them. From a practical
point of view, AC electrokinetic methods have the advantages of
short detection times and high sensitivity~\cite{ratanachoo02a}.

Dielectrophoresis can be defined as the movement of
polarizable particles in a non-uniform applied
AC electric field~\cite{Pohl}, whereas in
electrorotation an interaction between a
rotating AC electric field~\cite{Zim1} and suspended particles
leads to a rotational motion of the particles.
The most commonly used models to deal with the dielectric properties
of colloidal particles or biological cells are the so called shell models.
Because of inhomogeneous compartmentalization of biological
cells, one-, two-, and three- shell models have been applied to
discuss electrorotation of biological cells, e.g., see Refs.~\cite{Zim1,Fuhr2,Chan-2}.

These cell models have several limitations and they become complex
as the number of shells increases. This is particularly true when
two (or more) particles approach each other. In the dilute limit,
one can focus on the electrokinetic spectra of an individual
particle. If the suspension is not dilute, as it is often the case
in practice, the situation is complicated by the existence of
multipolar interactions. Even when a suspension is initially in
the dilute limit, particles often aggregate due to the presence of
an external electric field. In this case, a point dipole
approximation~\cite{Sauer1,Jones1990} becomes inadequate and the
mutual interactions must be taken into account by a
theory~\cite{PRE2,Huang:2003a} that goes beyond the point
dipole.

To provide a physically motivated and tractable model for inhomogeneous
particles, such as cells, we have recently
studied particles with spatial gradients in their
structures by introducing profiles for the conductivities
and dielectric constants of the particles, and
used differential approximation for the dielectric
factor~\cite{unsubmitted-1,huang03:b,Dong-PRB}.
Here, we extend this work to take into account polarization interactions
when two particles approach each other, and treat both DEP and ER using
the same theoretical framework.
We consider a pair of touching graded particles in suspensions.
As a result, the mutual polarization interactions and the
gradation fluctuations inside the particles lead to
significant changes in the electro-kinetic spectra.
In a more general context, these are manifestations of
correlation effects in charge-carrying system~\cite{grosberg:02a, patra03}.

\begin{figure}
\hspace*{-0.6cm}
\includegraphics[width=0.55\textwidth,clip=true,viewport=15 30 550 480]{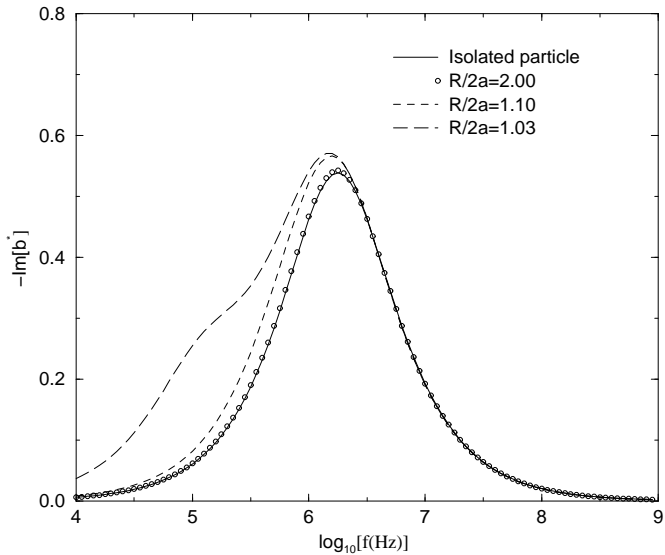}
\caption{ER spectra for an isolated particle and two touching
particles with $R/2a=2.00, 1.10, 1.03$, respectively. Parameters:
$c=-30\ep_0$, $m=1.0$.  } \label{Fig.2.}
\end{figure}

\section{Formalism}

We consider inhomogeneous biological cells or colloidal particles
with radius $a$. We assume that they have a distance-dependent complex dielectric
constant $\te_1(r)$ ($0<r\le a$), and that they are embedded in a host fluid having
dielectric constant
$\te_2$. Here $\te=\ep(r)+\sigma(r)/(2\pi i f)$, where $\ep$ denotes the
real dielectric constant, $\sigma(r)$ the conductivity, $f$ the
frequency of an external field, and $i\equiv \sqrt{-1}$. As the above formulas
suggest, $\ep(r)$ and $\sigma(r)$ are not constant inside the particle but
have distance dependent profiles. This is a very physical assumption and we will
return to it in the following discussion and  later in connection with the numerical simulations.

The dipole, or Clausius-Mossotti, factor reflects the polarization of a particle in a
surrounding medium.  In a recent work~\cite{unsubmitted-1}, we
derived the dipole factor for graded spherical particles by
introducing a differential effective dipole approximation (DEDA).
The generalization to the nonspherical case was done as
well~\cite{huang03:b}. The idea of the DEDA can be summarized as follows:
Consider a shell model for an inhomogeneous particle. In the DEDA one adds new
shells of infinitesimal thickness to the particle. Each of these
cells have distance-dependent complex dielectric constant. Since
the thickness of the layer approaches zero ($\mathrm{d}r
\rightarrow 0$), its correction to the dipole factor is
infinitesimal and one could eventually obtain a differential
equation.

The DEDA equation for a spherical graded particle has the
form~\cite{unsubmitted-1,huang03:b}
\begin{eqnarray}
\frac{{\rm d}b}{{\rm d}r} & = &  -\frac{1}{3r\te_2\te_1(r)}
              \left[(1+2b)\te_2-(1-b)\te_1(r)\right] \nonumber\\
\, & & \left[(1+2b)\te_2+2(1-b)\te_1(r) \right]. \label{DEDA}
\end{eqnarray}
It is worth noting that the DEDA is essentially exact since it
is in an excellent agreement with the exact solutions obtained
for a power-law profile and a linear profile by solving the
Laplace equation for the local electric field~\cite{Dong-PRB}.

For a pair of particles at a separation $R$ in a suspension, we
have to consider the multiple images effect~\cite{Yu00,PRE2}.
Theoretically, we may see the inhomogeneous graded
particle as an effectively homogeneous one. Then, we
consider two particles in a suspension which is subject to an
external uniform electric field. This yields a dipole moment into
each particle.  Let us denote the dipole moments of particles 1
and 2 as $p_{10}$ and $p_{20}$($\equiv p_{10}$ for identical particles),
respectively. Then,
we take into account the image effects. The dipole moment $p_{10}$
induces an image dipole $p_{11}$ into particle 2, while $p_{11}$
induces another image dipole in particle 1. As a result, multiple
images are formed. The same description holds for $p_{20}$.
Thus, we admit the infinite series of image dipoles. To this end,
we obtain the sum of dipole moments inside each particle, and
derive the desired expressions for dipole factors. Let us consider
two basic cases: 1) longitudinal field (L), where the field is
parallel to the line joining the centers of the particles, and 2)
transverse field (T), where the field is perpendicular to the line
joining the centers of the particles.

Based upon a multiple images method~\cite{Yu00},  the dipole
factors, $b_L{}^*$ and $b_T{}^*$, are given by~\cite{Yu00,PRE2}
\begin{eqnarray}
b_L^* &=& b\sum_{n=0}^{\infty}(2b)^n\left (\frac{\sinh\alpha}{\sinh(n+1)\alpha}\right )^3,\label{multi1}\\
b_T^* &=& b\sum_{n=0}^{\infty}(-b)^n\left
(\frac{\sinh\alpha}{\sinh(n+1)\alpha}\right )^3,\label{multi2}
\end{eqnarray}
for longitudinal and transverse field cases, respectively, where
$\alpha$ satisfies the relation $\cosh\alpha=R/2a$. 
Although it is not obvious, it is important to notice 
that multipoles are included in the above 
formulas~\cite{Sun+Yu:2003}. Clearly, the
multiple images effects have been taken into account in $b_L^*$
and $b_T^*$. It is worth noting that:  Setting $n$ up to $1$ in
the two equations leads to the dipole factor for two touching
particles in the point-dipole limit. In this case, in view of both
$|b|^2\ll 1$ and $R/2a\sim 1$, we have
\begin{eqnarray}
b_L^*(1) &=& \frac{b}{1-b/4},\nonumber\\
b_T^*(1) &=& \frac{b}{1+b/8}.\label{PD}
\end{eqnarray}
Both Eqs.(\ref{PD}) agree well with the result of Jones, which
were obtained by a field method in the point-dipole
limit~\cite{Jones1990}.

\subsection{Electrorotation}

By adding a rotating electric field with magnitude
$E_{\mathrm{ER}}{}^*$ to the two particle system, the effective
dipole factor for a pair of particles should be given
by~\cite{PRE2} \be b^*=(b_L{}^*+b_T{}^*)/2. \ee Thus, in this
case, the electrorotation velocity of a particle $\Omega^*$ is
given by~\cite{PRE2}
\be \Omega^*=-\phi
(\ep_2,\eta_2,E_{\mathrm{ER}}{}^*)\mathrm{Im}[b^*],
\ee where
$\phi (\ep_2,\eta_2,E_{\mathrm{ER}}^*)$ is a function of $\ep_2$,
the viscosity of the medium $\eta_2$, and $E_{\mathrm{ER}}{}^*$.
Here $\mathrm{Im}[\cdots]$ denotes the imaginary part of
$[\cdots]$. For an isolated spherical particle, $\phi
(\ep_2,\eta_2,E_{\mathrm{ER}}{}^*)=\ep_2E_{\mathrm{ER}}{}^*{}^2/2\eta_2$~\cite{Jones}.

\begin{figure}
\hspace*{-0.2cm}
\includegraphics[width=0.55\textwidth,clip=true,viewport=15 25 340 480]{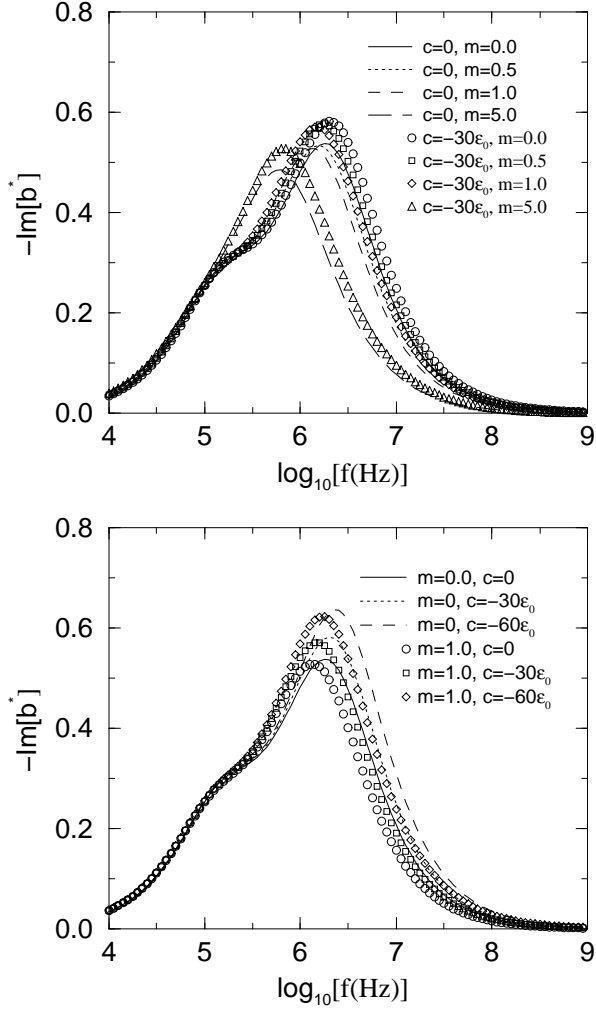}
\caption{ER spectra for two
touching particles. Upper panel for profile constants $m$ at $c=0$
and $c=-30\epsilon_0$, respectively. Lower panel for various $c$
at $m=0$ and $m=1.0$, respectively.  The spectrum is given as the
imaginary part of the dipole factor. Parameter: R/2a=1.03. }
\label{Fig.1.}
\end{figure}

\begin{figure}
\hspace*{-0.2cm}
\includegraphics[width=0.55\textwidth,clip=true,viewport=15 25 340 480]{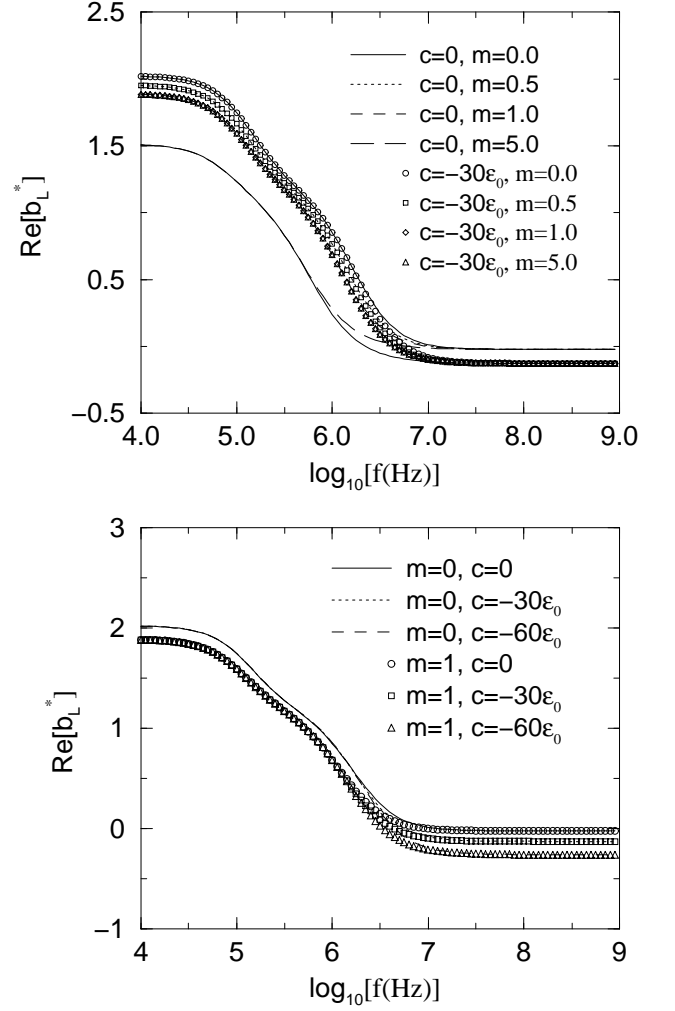}
\caption{DEP spectra for two touching particles in the
longitudinal field case. Upper panel for profile constants $m$ at
$c=0$ and $c=-30\epsilon_0$, respectively. Lower panel for various
$c$ at $m=0$ and $m=1.0$, respectively.  The spectrum is given as
the real part of the dipole factor. Parameter: R/2a=1.03. }
\label{Fig.4.}
\end{figure}

\subsection{Dielectrophoresis}

We consider a single particle suspended in a medium and subjected
to  a nonuniform AC electric field ${\bf E}_{\mathrm{DEP}}{}^*$.
The DEP force ${\bf F}_{\mathrm{DEP}}$ acting on the particle is
given by~\cite{Jones} \be {\bf F}_{\mathrm{DEP}}= 2\pi
\ep_2a^3\mathrm{Re}[b]\nabla |{\bf E}_{\mathrm{DEP}}{}^*|^2,
\label{depf} \ee where ${\bf E}_{\mathrm{DEP}}{}^*$ stands for the
local RMS electric field, and $\mathrm{Re}[\cdots]$ denotes the
real part of $[\cdots]$. Next, for a pair of touching particles,
the DEP force is given by~\cite{Huang:2003a,Jones}
\begin{eqnarray}
{\bf F}_L{}^* &=& 2\pi\ep_2a^3\mathrm{Re}[b_L{}^*]\nabla |{\bf E}_{\mathrm{DEP}}{}^*|^2,\\
{\bf F}_T{}^* &=& 2\pi\ep_2a^3\mathrm{Re}[b_T{}^*]\nabla |{\bf E}_{\mathrm{DEP}}{}^*|^2,
\end{eqnarray}
for longitudinal and transverse field cases, respectively.
The above formulation for the DEP force is, stricly speaking,
applicable for linearly polarized fields~\cite{Washizu+Jones:1996},
which is the case studied here.

\begin{figure}
\hspace*{-0.2cm}
\includegraphics[width=0.55\textwidth,clip=true,viewport=15 25 340 480]{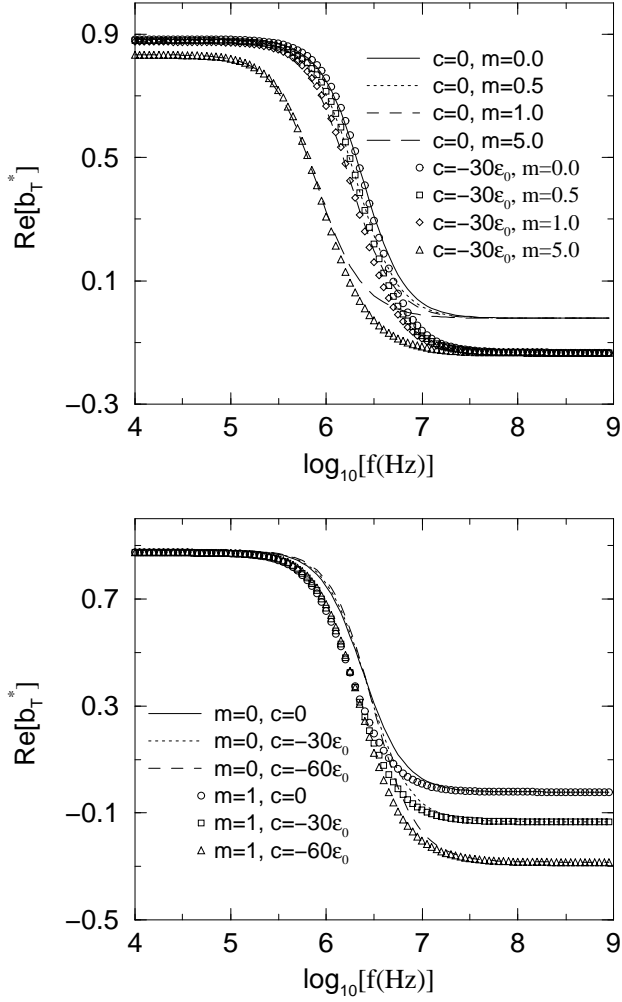}
\caption{Same as in Fig.~\ref{Fig.4.}, but in the transverse field
case.   } 
\label{Fig.5.}
\end{figure}

\begin{figure}
\hspace*{-0.2cm}
\includegraphics[width=0.55\textwidth,clip=true,viewport=15 25 340 480]{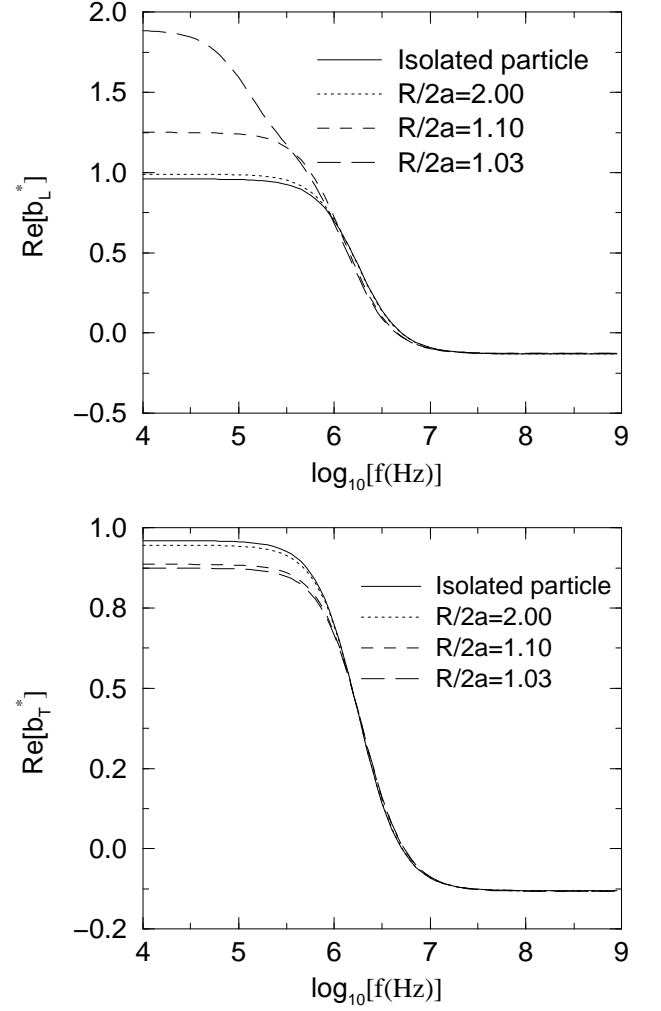}
\caption{DEP spectra in the longitudinal field case (upper panel)
and the transverse (lower panel), for an isolated particle, and
two touching particles with separation ratio $R/2a=2.00, 1.10,
1.03$, respectively. Parameters: $c=-30\ep_0, m=1.0$. }
\label{Fig.6.}
\end{figure}

\section{Numerical results}

For the following numerical calculations, we take the conductivity
and dielectric profiles to be
\begin{eqnarray}
\sigma_1(r)&=&\sigma_1(0) (r/a)^m,  \mathrm{\, \, \, \, \, \, \,} r \le a\\
\ep_1(r)&=&\ep_1(0)+c(r/a),         \mathrm{\, \,}                r \le a
\end{eqnarray}
where $m$ and $c$ are profile dependent constants. The profile is
clearly physical since conductivity can change rapidly
near the boundary of cell and a power-law profile
prevails~\cite{huang03:b}. On the other hand, the dielectric
constant may vary only slightly and thus a linear profile
suffices~\cite{huang03:b}.  In particular, the dielectric
constant at the center, namely $\ep_1(0)$, may be larger than that
at the boundary. Thus, in what follows, we would choose $c \le 0$.
By integrating the dielectric profile, we obtain an average
dielectric constant $\ep_{av}$ for different values of $c$ by
using a volume average~\cite{huang03:b}
\be
\ep_1{}^{av}=\frac{\int_0^a\ep_1(r)r^2{\rm d}r}{\int_0^ar^2{\rm d}r}.
\label{eave}
\ee
For the above dielectric profile
$\ep_1{}^{av}=\ep_1(0)+3c/4$.

For all numerical calculations,  we take
$\epsilon_1(0)=75\epsilon_0$, $\epsilon_2=80\epsilon_0$,
$\sigma_1(0)=2.8\times 10^{-2}$S/m, and  $\sigma_2=2.8\times
10^{-4}$S/m. Here $\epsilon_0$ denotes the dielectric constant of
the free space.

Figure~\ref{Fig.2.} shows the ER spectrum of two particles at different distances
from each other. At large separations (e.g. $R/2a>2$), the multipolar interaction may be
neglected, but the induced multiple images play an
important role in the spectrum when two particles approach each other.

In Fig.~\ref{Fig.1.}, it is evident that a second peak due to the
multiple image effect occurs at a lower frequency. In fact, the
appearance of a second peak has been predicted for homogeneous
particles in a recent work~\cite{PRE2}.  Moreover, fluctuations
in the conductivity profile can make the characteristic frequency
shifted to lower frequencies
(red-shifted), while those in the dielectric profile can enhance the
peak value. However, such effects on the second characteristic
frequency and its peak value are small enough to be neglected.

Fluctuations in conductivity and dielectric profiles may enhance
the DEP spectrum not only in the longitudinal field case (see
Fig.\ref{Fig.4.}), but also in the transverse (see
Fig.\ref{Fig.5.}). The effects of multiple images may change the
DEP spectrum significantly.

Similar to Fig.~\ref{Fig.2.}, Figure~\ref{Fig.6.} shows that the
multiple images play a crucial role in the DEP spectrum when the
particles are close enough. In contrast to the result from an
isolated particle, the multiple image effect may enhance the DEP
spectra (moreover, the real part of the dipole factor may be
enhanced to be larger than $1$) at low frequency region in the
longitudinal field case. However, the DEP spectrum is reduced in
the transverse field case due to the presence of multiple images.

In addition, we have also compared the
point dipole model with the current multiple image dipole model
(no figures shown). As expected, the results predicted
by them are quite different, especially at low frequencies. This
shows further that the point dipole model is inadequate for the
touching particles, and thus needs to be modified to take into account
the effect of multiple images.

\section{Discussion and conclusion}

In this work, we investigated the effects of multipolar
interactions on AC electrokinetic behavior, electrorotation and
dielectrophoresis, of inhomogeneous biological cells and colloidal
particles. We model such inhomogeneous particles as graded ones. Our
method may be extended to high concentration
case~\cite{Huang:2002a,Gao-1} or pearl chain case~\cite{PRE1,JAP1},
work is in progress to address these issues in detail. Also,
it is possible to take into account shape effects by considering the
nonspherical shapes, such as oblate or prolate
spheroid~\cite{JPCM1}. In doing so, we might resort to the derived
DEDA equation for graded spheroidal particles as
well~\cite{huang03:b}. 
It is also straightforward to extend this
work to deal with the experimentally interesting case of
charged colloidal suspensions~\cite{Huang:2003a}.

To put our approach in the context of composite particles, 
we have performed a mean-field approach in the spirit of 
Choy {\em et al.}~\cite{Choy:1998a,Choy:1998b}, i.e., treating  inhomogeneous
particles as effectively homogeneous ones which are embedded in a
uniform field. In particular, it is worth noting that well-known Tartar 
formula~\cite{Milton:2002}  can be used to exactly calculate the effective complex
dielectric constant of a single graded particle.  Thus, once this effective
complex dielectric constant is obtained, one can proceed to
calculate the relevant dipole moment, and hence the desired dipole factor.
More interestingly, our DEDA  approach can predict exactly the same result as Tartar
formula~\cite{gao-repr}.

To sum up, based on the DEDA, we have presented a theoretical study of
electrokinetic behavior, electrorotation (ER)  and
dielectrophoresis (DEP) for two touching inhomogeneous particles
in suspensions.  We found mutual polarization effects and the
spatial fluctuations inside colloidal particles or biological
cells can both affect ER and DEP spectra significantly. 
Our approach
has the further advantage of being able to treat both
electrorotation and dielectrophoresis using the same theoretical framework.

As a further study and a test to our theory, it would be interesting
to have a systematic experimental investigations of these effects. 
The hope is that they would shed light to the limits of the theory and that
they would help to separate the DEP and ER behavior from, e.g. electrohydrodynamic
flow effects~\cite{Voldman:2001aa} and limitations due to Brownian motion.  
One possibility for doing so would be to use the laser tweezers
combined with ER and/or DEP~\cite{Schnelle:2000aa,Voldman:2001aa}.

\acknowledgments

This work has been supported by the Research Grants Council of the Hong
Kong SAR Government under project number CUHK 4245/01P, and by the
Academy of Finland (M.\,K.) and the Finnish Academy of Science
and Letters (M.\,K.). 
One of us (M.\,K.) would like to thank Andreas Manz for inspiring discussions.

\end{document}